# Cooper Pair Writing at the LaAlO$_3$/SrTiO$_3$ Interface

Cheng Cen[1], Daniela F. Bogorin[1], Chung Wung Bark[2], Chad M. Folkman[2], Chang-Beom Eom[2], Jeremy Levy[1*]

[1]Department of Physics and Astronomy, University of Pittsburgh, 3941 O'Hara St., Pittsburgh, PA 15260, USA.

[2]Department of Materials Science and Engineering, University of Wisconsin-Madison, Madison, Wisconsin 53706, USA

PACS numbers: 73.20.-r  74.78.-w  74.72.Kf

The LaAlO$_3$/SrTiO$_3$ interface provides a unique platform for controlling the electronic properties of the superconducting semiconductor SrTiO$_3$. Prior investigations have shown that two-dimensional superconductivity can be produced at the LaAlO$_3$/SrTiO$_3$ interface and tuned electrostatically. The recently demonstrated reversible control of the metal-insulator transition at the same interface using conductive atomic force microscopy (c-AFM) raises the question of whether this room-temperature technique can produce structures that exhibit superconducting, normal metallic and insulating phases at sub-Kelvin temperatures. Here we report low-temperature magnetotransport experiments on conducting structures defined at an otherwise insulating LaAlO$_3$/SrTiO$_3$ interface. A quantum phase transition associated with the formation of Cooper pairs is observed in these predefined structures at sub-Kelvin temperatures. However, a finite resistance remains even at the lowest temperature. At higher magnetic fields, interfaces with high mobility also exhibit strong Shubnikov-de Haas oscillations as well as a larger Ginsburg-Landau coherence length. Cooper pair localization, spin-orbit coupling, and finite-size effects may factor into an explanation for some of the unusual properties observed.

[*]To whom correspondence should be addressed.  E-mail: jlevy@pitt.edu

Electrostatic control over superconducting (S), normal metallic (N) and insulating (I) phases [1, 2]—at scales comparable to superconducting coherence length—is highly desired in making superconducting electronics. The ability to define superconducting regions with fully insulating backgrounds is a prerequisite for superconducting electronics, and forms the basis for creating Josephson junctions, superconducting quantum interference devices (SQUIDs), resonator structures[3] and Cooper pair boxes [4]. Analog-style control over carrier density is important for creating normal-superconducting junctions that interconvert Cooper pairs and normal-state quasiparticles, which can be used to create spin-entangled electron pairs [5].

The interface between a thin $LaAlO_3$ film (thickness of $d_{LAO}$) and $TiO_2$-terminated bulk $SrTiO_3$ undergoes a sharp phase transition from an insulating state for $d_{LAO} \leq 3$ unit cells (uc) to a conducting state for $d_{LAO} \geq 4$ uc [6]. At low temperatures, superconductivity has been reported [7] with transition temperatures and equivalent three-dimensional carrier densities comparable to those first reported by Schooley et al [8]. Macroscopic control of the superconducting phase at the $LaAlO_3/SrTiO_3$ interface was achieved by electrostatic tuning from the bottom of the $SrTiO_3$ substrate [9]. However, controversy does exist in the origin of the interface conductance. Especially in planar as grown structures, complete exclusion of either electron doping or chemical doping is difficult to obtain [10-14]. Using a conductive atomic force microscope (c-AFM) writing technique, conducting areas can be produced at an initially insulating $LaAlO_3/SrTiO_3$ interface [15-20]. The writing process is a form of modulation doping in the sense that localized positive charges at the top $LaAlO_3$ surface create electronic states at the $LaAlO_3/SrTiO_3$ interface. This allows for transport studies to be conducted where the electrons are produced entirely by these remote dopants.

Here we report low-temperature magnetotransport measurements on conducting channels created at the LaAlO$_3$/SrTiO$_3$ interface using c-AFM [18]. The heterostructures consist of 3uc LaAlO$_3$ films grown on TiO$_2$-terminated (001) SrTiO$_3$ substrates by pulsed laser deposition (see supporting online materials for detailed material fabrication method). Two types of heterostructures are fabricated. "Type-H" structures are grown under an oxygen pressure of 10$^{-3}$ mbar at 550ºC while "Type-L" structures are grown under an oxygen pressure of 7.5 × 10$^{-5}$ mbar at 780ºC. Several samples of each type are fabricated and measured. For consistency, data is shown only for one sample per type (H1 and L1).

Utilizing c-AFM lithography (Fig. 1(a)), conducting and insulation regions are defined at the interface of LaAlO$_3$/SrTiO$_3$ [18]. Hall bar structures composed of one 1 μm wide current channel and two 1 μm wide voltage sensing channels are written at room temperature between Au electrodes which are in ohmic contact with the interface (Fig. S1 [21]). The structures are written at room temperature with an AFM tip voltage $V_{\text{tip}}$=+10 V and subsequently cooled using a dilution refrigerator.

Magnetotransport measurements carried out at 20 mK on Hall bar structures created with same writing parameters on several samples of both H type and L type all result in sheet carrier density $n_s$~4×10$^{13}$ cm$^{-2}$. Measurements of sheet resistances result in a Hall mobility 1400 cm$^2$V$^{-1}$s$^{-1}$ for sample H1 and 160 cm$^2$V$^{-1}$s$^{-1}$ for sample L1. In a field range between 3T and 8T, the magnetoresistance under perpendicular magnetic fields in sample H1 exhibits Shubnikov-de Haas (SdH) oscillations (Fig.2(a)), consistent with the higher carrier mobility. A close examination of the oscillations by extracting a positive magnetoresistance background reveals a clear periodic pattern as a function of the reciprocal of field strength whose amplitude decays at elevated temperatures (Fig. 2(b)). A Fourier transform (Fig. 2(c)) reveals a strong frequency peak at 31 T. Assuming a two-dimensional free electron model with spin degeneracy, this frequency corresponds to a sheet carrier density $n_{SdH} = 1.5\times10^{12}$ cm$^{-2}$. A multi-subband structure of SrTiO$_3$ may be responsible for the discrepancy between carrier density measured by SdH effect and that from Hall measurements [22, 23], and has been observed as well in delta-doped SrTiO$_3$ layers [24]. SdH oscillations vanish when magnetic field becomes parallel with the sample surface, evidencing a two dimensional transport system. SdH oscillations are not observed in L-type samples.

At zero magnetic field, sample H1 undergoes a phase transition below a temperature $T_P^{H1}=$ 200 mK in which the resistance decreases by 2% (Fig.1(b) and Fig.3(a)). Application of a plane-perpendicular magnetic field greater than a critical value $H_{C2\perp}^{H1}$ ~4 mT completely suppresses this transition (Fig.1(b) and Fig.3(a-b)). A similar transition is observed in L1 within the measurable temperature range and with a higher transition temperature $T_P^{L1} > 370$ mK (Fig.3(a)). Sample L1 also exhibits a more than one order of magnitude larger critical field $H_{C2\perp}^{L1}$ ~70 mT (Fig.3(b)). Samples H1 and L1 both exhibit a critical current density (Fig.3(c)), above which the resistance returns to the normal-state value (i.e., the value above $T_P$ or $H_{C2\perp}$).

Given that LaAlO$_3$/SrTiO$_3$ heterostructures with thicker LaAlO$_3$ films ($d_{LAO} \geq 4$ uc) are known to undergo a superconducting phase transition in similar temperature range [7, 9], it is reasonable to relate this behavior to superconductivity. Sheet resistances of the conducting channels in H1 and L1 are both well below the universal value $h/4e^2 \approx 6.5$ kΩ/□. Therefore, the systems meet one general condition for the attractive interactions between electron pairs to overcome localization effects of each electron. The destruction of the lower resistance state of both H1 and L1 by a sufficiently large magnetic field closely resembles the Meissner effect. The existence of a critical current is also a typical signature of superconducting state. However, the presence of a zero-resistance state, the hallmark of superconductivity, is strikingly absent. Before further discussion of the lack of a zero-resistance state, we first describe more observations with a preliminary analysis that utilizes the Ginsburg-Landau (GL) phenomenological framework of superconductivity.

The upper critical field $H_{C2\perp}(T)$ in both H1 and L1 decreases with an increasing temperature. Data for structure H1 can be fit to a GL formula:

$$H_{C2\perp}(T) = \frac{\phi_0}{2\pi\xi_{GL}(0)^2}(1 - \frac{T}{T_C}) \qquad (1)$$

where $\phi_0$ is the flux quantum and $\xi_{GL}(0)$ is the GL coherence length at zero temperature (Fig.4(a)). From the fitting, a coherence length $\xi_{GL}^{H1}(0) = 240$ nm can be extracted. Due to the uncertainty of $T_P^{L1}$ and fewer temperature dependent data points available, a reliable fit is not obtained for sample L1 (Fig.4(a)), but from the data well below $T_P^{L1}$, a coherence length $\xi_{GL}^{L1}(0) = 70$ nm can be estimated.

Characteristics of the Meissner effect vary significantly with magnetic field orientation (Fig.4(b)). The change of $H_{C2}$ as a function of the angle $\theta$ between the sample surface and the magnetic field agrees well with the two-dimensional functional form [25]:

$$\left| \frac{H_{C2}(\theta)\sin\theta}{H_{C2\perp}} \right| + \left( \frac{H_{C2}(\theta)\cos\theta}{H_{C2\parallel}} \right)^2 = 1. \tag{2}$$

For both sample H1 and L1, $H_{C2}$ increases by at least one order of magnitude when the magnetic field is parallel to the interface ($H_{C2\parallel}^{H1} = 50$mT; $H_{C2\parallel}^{L1} = 890$mT), indicating that the transport layer $d$ is much thinner than the coherence length. The measured $H_{C2\parallel}^{H1}$ is expected to be smaller than the actual value due to the presence of an off-axis stray magnetic field ($H_{stray} \sim $ 1mT, comparable to $H_{C2\perp}^{H1}$ =4mT) in the superconducting magnet (see supplementary materials [21]). The transport layer thicknesses for H1 and L1 can be estimated with [25]:

$$H_{C2\parallel}(T) = \frac{\phi_0 \sqrt{12}}{2\pi \xi_{GL}(0) d} (1 - \frac{T}{T_C})^{1/2} \tag{3}$$

Calculations based on this form yield $d^{H1}$ = 87 nm and $d^{L1}$ = 18 nm, however $d^{H1}$ is overestimated because of the stray field affected $H_{C2\parallel}^{H1}$ measurement value. Equivalent three dimensional carrier densities can be calculated: $n_{3D}^{H1} \approx 5 \times 10^{18}$ cm$^{-3}$, $n_{3D}^{L1} \approx 2 \times 10^{19}$ cm$^{-3}$. The smallest carrier density for bulk SrTiO$_3$ to be superconducting with $T_C$ > 200 mK is $3 \times 10^{19}$cm$^{-3}$ [26]. The calculated values for $n_{3D}^{H1}$ and $n_{3D}^{L1}$ clearly fall below this lower limit.

In all the experiments discussed above, no voltage bias is applied to the back of SrTiO$_3$ substrate. In L-type samples, biases applied to the back of SrTiO$_3$ substrate can tune the transport properties of the interface conductive structures as other reported results taken on samples grown under the comparable conditions [6, 9]. However, transport in H-type samples is not measurably affected by the back biases. This distinction is consistent with the previous recognition that the field effect on interface transport induced by back biases involves the movement of oxygen vacancies in SrTiO$_3$ substrate [6], which are lacking in well-oxidized type-H samples.

Back biases are later applied to the L-type samples. The zero-resistance state remains absent with positive biases $V_{back}$ up to 200V. After sweeping the back bias down to zero, all channels become insulating in L1 and a back bias larger than 60V is needed to restore the conductance within the measurable range thereafter. Similar effects were also reported elsewhere and were explained by irreversible filling of trap states when interface electrons are pulled deeper into SrTiO$_3$ substrate by large positive back bias [27]. Varying $V_{Back}$ from 60V to 80V, the overall channel resistance drops almost one order of magnitude (Fig. 5). The magnetic field induced resistance change $\Delta R=R(100mT)-R(0T)$ and $H_{C2}$ remain much less affected but are both considerably smaller than the value measured before applying any back bias (Fig.5). We note that the effect of a back bias is expected to differ from those previously reported [9, 27], since the electrical boundary conditions are significantly different for a 1μm channel surrounded by an insulating background, rather than for a planar conducting (or superconducting) interface. Factors such as the vertical and in-plane redistribution of the electrons originally confined in interfacial structures need to be considered in order to fully account for the back gating effect.

Except for the absence of a zero-resistance state, the written LaAlO$_3$/SrTiO$_3$ interfacial structures exhibit properties consistent with Cooper pair formation and superconductivity described by Ginzburg-Landau framework. In system with low carrier density, the effect of phase fluctuations becomes significant [28]. Both intrinsic quantum fluctuations (i.e., the Berezinskii-Kosterlitz-Thouless quantum phase transition) [7, 28-31] and fluctuations arising from extrinsic sources of disorder can destroy superconducting long range order. In the situations where the coherence length becomes comparable to the order length scale, a fully superconducting state may cease to exist. In SrTiO$_3$ (especially thin films), a regime exists in which Cooper pairs form but are not fully condensated to a single coherent state [32]. Similar effects has also been observed in cuprate superconductors, where Cooper pairs are present within a pseudogap above the superconducting transition temperature [33]. Considering the low equivalent three-dimensional carrier densities, the lower resistance states we observed may be interpreted as such a pairing state when the phase transition temperatures $T_P^{H1}$ and $T_P^{L1}$ represent the corresponding pairing temperature.

Other mechanisms may as well contribute to the suppression of a truly superconducting state, such as larger-scale inhomogeneities and finite-size effects. Inhomogeneity in the conducting channels caused by local defects and disorder may also contribute to a spatially nonuniform transition, in which some sections of the channels remain in a normal state even at the lowest temperature measured, giving rise to a metallic series resistance. However, in the case of an inhomogeneous channel, a graduate reduction of resistance will most likely be present instead of a well-defined transition temperature observed in our experiment [34]. The 1μm in-plane confinement width is larger but still comparable with the in plane Ginzburg-Landau coherence lengths. Such lateral confinement may produce phase slips and thereby losses as observed in other low-dimensional superconductors [35]. We erased the initial Hall bar structure in sample L1 and rewrote a 50 μm × 50 μm conducting square at the same electrode set (Fig. S3(a) [21]). A sharp ~5% resistance drop (comparable to the resistance change in the initial Hall bar structure) takes place at 350 mK. However, a finite resistance still maintains even to the lowest temperature measured (Fig. S3(b) [21]). With the extended structure size, effects of lateral confinement and inhomogeneity are supposed to be largely suppressed. The fact that a similar transition to a lower but nonzero resistance state is observed indicates that mechanisms other these effects (such as low carrier density) may be responsible to the residual resistance.

Finally, we note that strong spin-orbit coupling is known to be present in this system [36]. Because nearly all of the donors originate from the top LaAlO$_3$ surface, we expect even stronger Rashba spin-orbit interaction (RSOI) strengths compared to structures in which oxygen vacancies contribute significantly to the interfacial electron gas. A sufficiently strong RSOI on a superconducting state can lead to complete suppression of the quasiparticle gap and the emergence of a topological superconducting state that is possibly useful for quantum computation [37].

Future exploration of AFM written nanowires which inherit the properties of the LaAlO$_3$/SrTiO$_3$ interface may enable a further simplified experimental realization of such a topological superconductor. The ability to define regions that support Cooper pair formation marks an important first step toward that goal.

In conclusion, conductive structures are fabricated using AFM lithography technique at LaAlO$_3$/SrTiO$_3$ interfaces grown under different conditions. Evidence for Cooper pairing of electrons is observed through transport experiments performed as a function of temperature and magnetic field. With future experiments, improved coherence of Cooper pairs may be obtained via controlling parameters like writing conditions and oxygen vacancies. The ability to control—with high spatial precision—electron pairing surrounded by fully insulating background opens a plethora of opportunities making on-demand Cooper pair based devices and circuits.

**Acknowledgement**: Work at NHMFL is performed under the auspices of the National Science Foundation, Department of Energy and State of Florida. Work at the University of Wisconsin was supported by funding from the National Science Foundation (DMR-0906443) and the DOE Office of Basic Energy Sciences under award number DE-FG02-06ER46327. Work at the University of Pittsburgh was supported by the National Science Foundation (DMR-0704022) and the Fine Foundation. We thank Ju-Hyun Park at NHMFL for experimental assistance. We also thank Harold Y. Hwang and Chetan Nayak for helpful discussions.

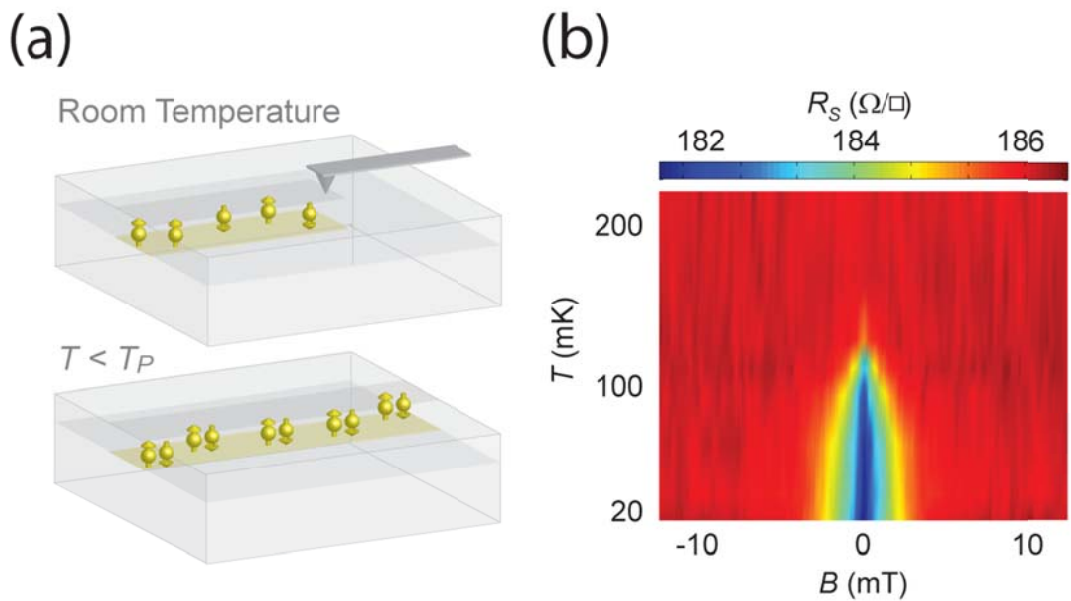

**Figure 1** (a) Conducting channels are written with c-AFM at room temperature and then cooled in dilution refrigerator. Below temperature $T_P$, a quantum phase transition is observed which is associating with the formation of cooper pairs. (b) Sheet resistance of the channel in sample H1 plotted versus the plane-perpendicular magnetic field and temperature, showing clear separation of normal phase and a lower resistance phase.

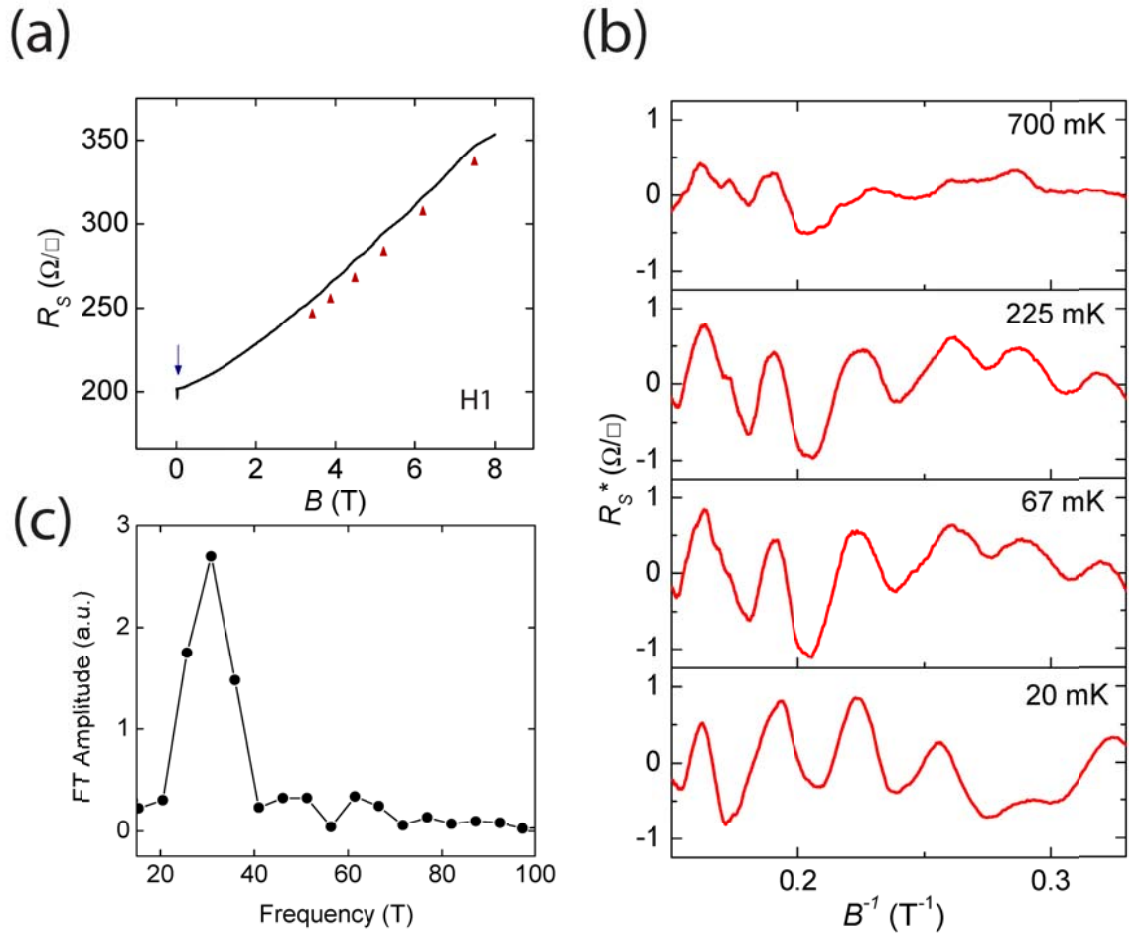

**Figure 2** Shubnikov-de Haas oscillations observed in sample H. (a) Magnetoresistance measured in sample H at 20 mK. Blue arrow points to the superconducting transition observed at low field. Red triangles points to SdH oscillations measured at high field. (b) SdH oscillations measured at elevated temperatures. A polynomial fit of the magnetoresistance is subtracted to reveal the oscillations more clearly. (c) Fourier transform (FT) of the SdH oscillations at 20 mK with an amplitude peak at 31 T.

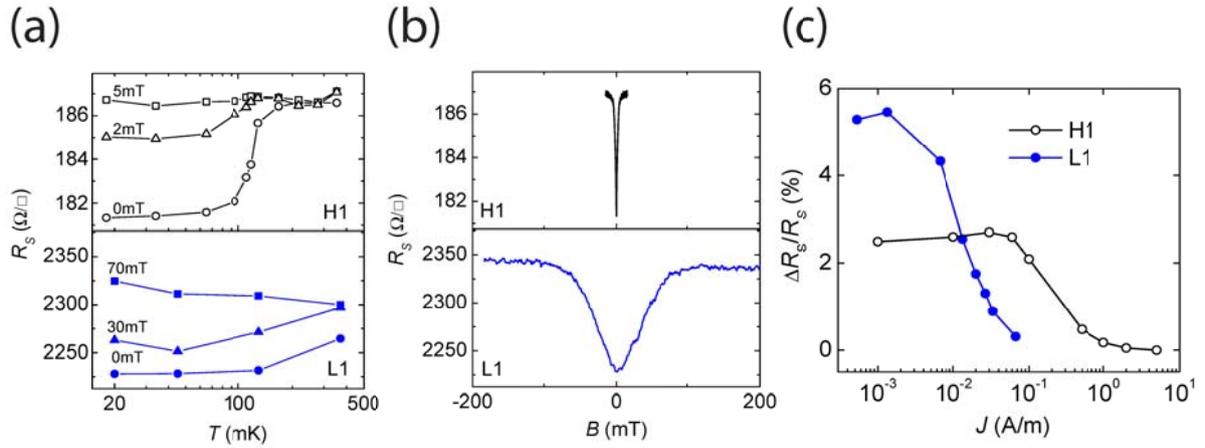

**Figure 3** Superconducting transition observed in interfacial conducting structures. (a) Sheet resistance of channel $R_S$ under different magnetic fields plot as a function of temperature, showing a transition temperature of 196mK for sample H1 and in excess of 370 mK for sample L1. (b) Suppression of the lower resistance state by applying a plane-perpendicular magnetic field. Sample L1 has a significantly larger upper critical than sample H1. (c) Percentage change of channel resistance between superconducting state and normal state plotted as a function of current density.

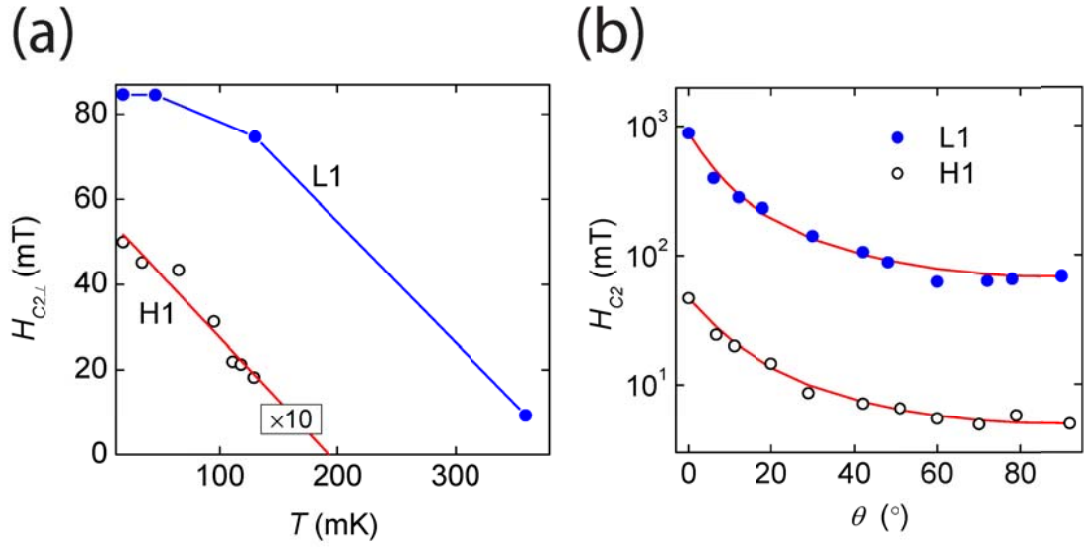

**Figure 4** Superconducting characteristics of samples (a) Upper critical field $H_{C2\perp}$ measured at various temperatures when magnetic field is perpendicular with sample surface in L1 (filled circle, value multiplied by 0.1 for clarity) and H (open circle). Red line is fit to Landau-Ginzburg theory. (b) $H_{C2}$ measured at 20 mK with different magnetic field orientations from 0º (field parallel with sample surface) to 90º (field perpendicular with sample surface). Red lines are fits to equation (2).

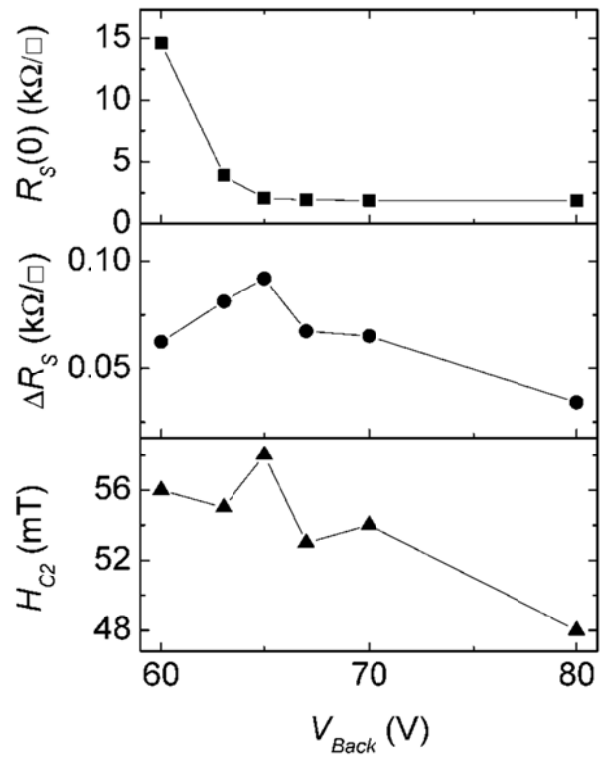

**Figure 5** Change of zero field channel resistance $R_S(0)$, magnetic field induced resistance change $\Delta R_S$ and upper critical field $H_{C2}$ induced by the back bias.

# Cooper Pair Writing at the LaAlO$_3$/SrTiO$_3$ Interface: Supplementary Materials

Cheng Cen[†], Daniela F. Bogorin[†], Chung Wung Bark[‡], Chad M. Folkman[‡], Chang-Beom Eom[‡], Jeremy Levy[†*]

[†]Department of Physics and Astronomy, University of Pittsburgh, 3941 O'Hara St., Pittsburgh, PA 15260, USA.

[‡]Department of Materials Science and Engineering, University of Wisconsin-Madison, Madison, Wisconsin 53706, USA

**Material Fabrication Method**

The LaAlO$_3$ thin films were deposited on TiO$_2$ terminated (001) SrTiO$_3$ substrates via pulsed laser deposition (PLD) with *in situ* high pressure reflection high energy electron diffraction (RHEED)[1] using two slightly different recipes. Samples of L type were deposited at 780°C at 7.5×10$^{-5}$ mBar O$_2$ pressure, then cooled down to atmospheric pressure with a one hour annealing at 400mBar O$_2$ pressure. Samples of H type were deposited at 550°C at 10$^{-3}$ mBar O$_2$ pressure. The difference of oxygen pressure during the deposition created samples with different concentrations of vacancies in the SrTiO$_3$ towards the 2DEG interface. Subsequently the "H-Type" sample will not be back gateable while "L-Type" samples will respond to an applied back gate bias; see Figure 5.

Ohmic contacts to the 2DEG interface between LaAlO$_3$/SrTiO$_3$ are obtained by sputtering electrodes as follow. The samples have been patterned via optical lithography; a 25nm deep area in the LaAlO$_3$/SrTiO$_3$ layer has been milled away using an Ar$^+$ Ion mill and backfilled with 2nm thick Ti for good adhesion to SrTiO$_3$ and 23nm Au electrodes. Hall bars are further defined between Au/Ti electrodes by raster scanning an conducting atomic force microscope (c-AFM) at speed of ~1um/s.

**Structure layout**

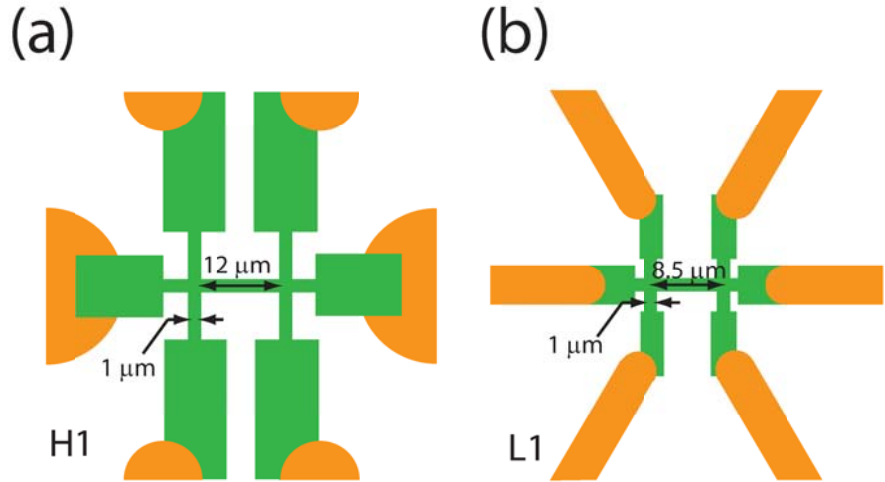

**Figure S1** Layout of structures written at the LaAlO$_3$/SrTiO$_3$ interface in sample H1 (a) and L1 (b). Orange areas represent Au electrodes; green areas indicate conducting channels written with c-AFM.

**Stray magnetic field**

In the presence of an off-axis stray magnetic field $B_{stray}$, magnetic field component perpendicular to the sample surface can be expressed as $B_\perp = B\sin\theta + B_{stray}\cos\theta$, where $B$ is sweeping magnetic field parallel with the magnet axis and $\theta$ is the angle between the sweeping field and the sample. At nonzero $\theta$, the net $B_\perp$ can be minimized at $B = -B_{stray}\tan^{-1}\theta$. When the sample is rotated parallel to the magnet axis ($\theta = 0º$), $B_\perp = B_{stray} \approx 1$mT and can no longer be minimized by the sweeping magnet field $B$. In sample H-1, the stray field is comparable to the upper critical field, giving rise to a clear increase of the minimum resistance at $\theta = 0°$ (Fig. S1A), while in sample L-1, due to the much larger upper critical filed, such effect is not obvious (Fig. S1B).

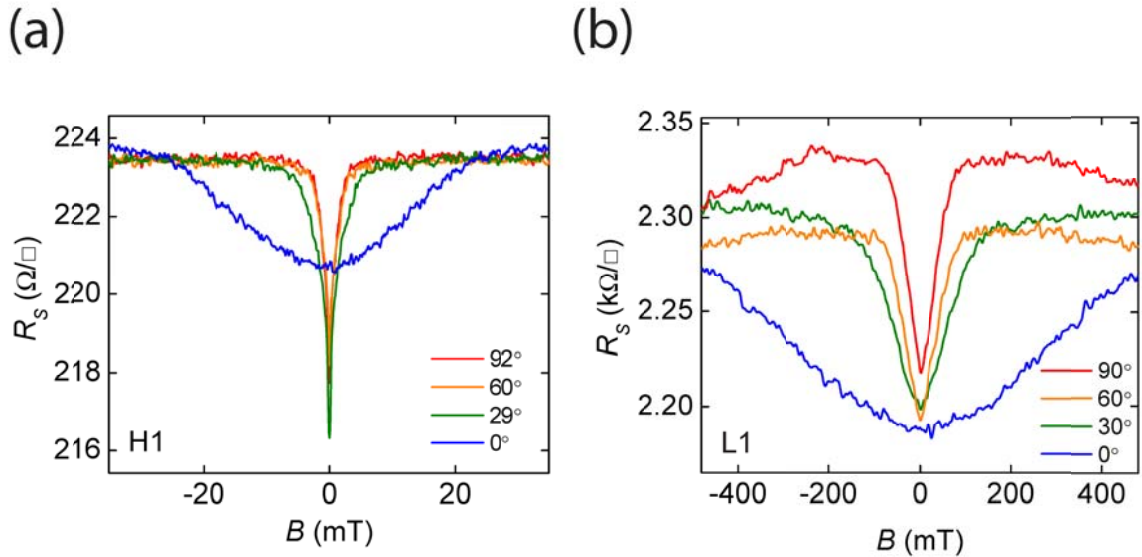

**Figure S2** Meissner effect measured when sample tilted with different angles relative to the superconducting magnet axis. (a) When sample H1 is parallel with the magnet axis, the minimum resistance level of the channel increases due to the small $H_{C2}$ (4mT) and an off-axis stray field (~1mT) in the superconducting magnet that cannot be compensated in this orientation. (b) For sample L1, since $H_{C2}$ (70mT) is much larger than the off-axis residual field, the large change in minimum resistance is not observed.

**Phase transition in larger structures**

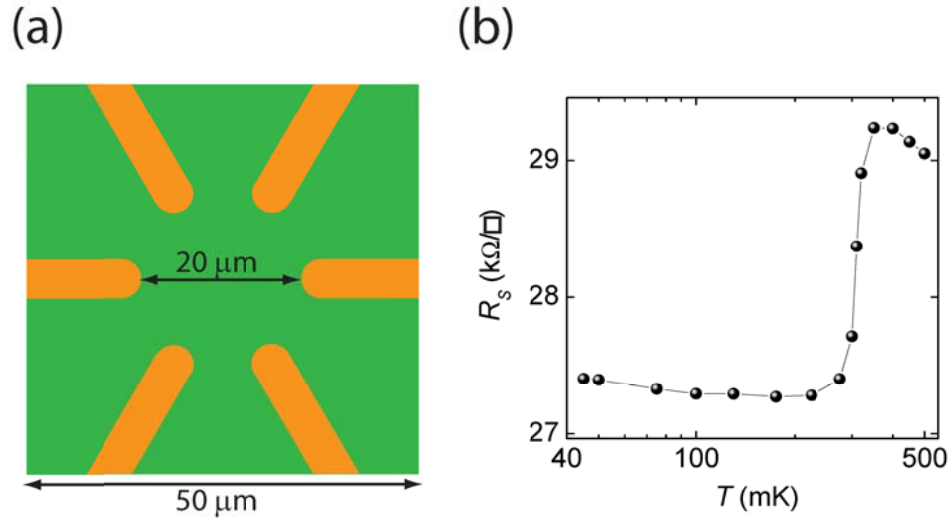

**Figure S3** (a) A 50 μm × 50 μm conducting square area is written at the same electrode set after the erasure of the Hall bar in L1. (B) Sheet resistance calculated from four probe resistance measurement using finite element simulation assuming a uniform conductivity. Transition to a lower but still finite resistance state is observed at 350 mK.

**References**

1. Rijnders, G. J. H., Koster, G., Blank, D. H. A. & Rogalla, H. *In situ* monitoring during pulsed laser deposition of complex oxides using refection high energy electron diffraction under high oxygen pressure. *Appl. Phys. Lett.* **70**, 1888-1890 (1997).